\documentclass{article}
\usepackage{spconf,amsmath,graphicx,amsfonts}
\usepackage{algorithm}
\usepackage{algorithmic}
\usepackage{amssymb}

\newtheorem{assumption}{\textbf{Assumption}}
\newtheorem{theorem}{\textbf{Theorem}}
\title{Decentralizing Coherent Joint Transmission Precoding via Deterministic Equivalents}

\name{\normalsize{Yuhao Liu$^{\ast\ddagger}$ \thanks{$^{\ast}$Equal Contribution}, Xinyu Bian$^{\ast\dagger\diamond}$, Yizhou Xu$^{\ddagger}$, Tianqi Hou$^{\diamond}$, Wenjie Wang$^{\diamond}$, Yuyi Mao$^{\S}$, and Jun Zhang$^{\dagger}$}}
\address{\normalsize{$^{\ddagger}$Dept. of Mathematical Sciences, Tsinghua University, Beijing, China}\\
\normalsize{$^{\dagger}$Dept. of ECE, The Hong Kong University of Science and Technology, Hong Kong, China}\\
\normalsize{$^{\diamond}$Theory Lab, Central Research Institute, 2012 Labs, Huawei Technologies Co., Ltd., Hong Kong, China}\\
\normalsize{$^{\S}$Dept. of EEE, The Hong Kong Polytechnic University, Hong Kong, China}}
\begin{document}
%
\maketitle
\begin{abstract}
In order to control the inter-cell interference for a  multi-cell multi-user multiple-input multiple-output network, we consider the precoder design for coordinated multi-point with downlink coherent joint transmission. To avoid costly information exchange among the cooperating base stations in a centralized precoding scheme, we propose a decentralized one by considering the power minimization problem. By approximating the inter-cell interference using the deterministic equivalents, this problem is decoupled to sub-problems which are solved in a decentralized manner at different base stations. Simulation results demonstrate the effectiveness of our proposed decentralized precoding scheme, where only $2 \sim 7\%$ more transmit power is needed compared with the optimal centralized precoder.
\end{abstract}
\section{Introduction}\label{sec:intro}
Attributed to the boost of advanced mobile applications, global mobile data traffic is increasing dramatically, which poses great challenges to cellular networks in providing high throughput and seamless coverage. In order to satisfy the demand of mobile broadband, network densification is considered as a promising solution \cite{sca2000}. However, the densely deployed base stations (BSs) bring severe inter-cell interference, which significantly affects the quality of service (QoS), especially for the cell edge users \cite{xyou2011}. To overcome this problem, coordinated multi-point (CoMP) with coherent joint transmission (CJT) was proposed \cite{msa2010}, where a user equipment (UE) may be served by several BSs simultaneously to mitigate inter-cell interference. It in turn increases the desired signal power, and hence improves spectral efficiency. In particular, precoding is one of the most fundamental approach to suppress interference among CoMP cells, which can control and even completely eliminate the inter-user and inter-cell interference \cite{fra1998}.

Many approaches on the precoder design for CJT have been proposed. Specifically, non-linear dirty-paper coding (DPC) was proposed for multi-cell multi-user multiple-input multiple output (MIMO) in \cite{hzha2004}. However, DPC requires complicated nonlinear encoding and decoding schemes and is difficult to implement. Therefore, linear beamforming schemes have been exploited for broadcast MIMO systems, such as zero forcing (ZF) \cite{ssh2008}, where the inter-user interference can be eliminated and each user perceives an interference-free MIMO channel. Based on the ZF method, Zhang et al. \cite{jzhang2009} further proposed a clustered base station coordination strategy for the downlink of a large cellular MIMO network, which can mitigate interference for both cluster interior and cluster edge users while reducing the complexity and channel state information requirements. Furthermore, considering the power constraints  of each antenna at the BS, Wu and Fei \cite{zwu2018} proposed a method based on the weighted minimum mean square error (WMMSE) approach \cite{qshi2011}, which aims at maximizing the sum rate of all users. However, these methods require the instantaneous channel state information (CSI) from all BSs to calculate the precoding matrices in a central unit, and then distribute the precoding matrices to BSs, which results in heavy signaling overhead and latency, especially when the number of cooperating BSs in the network becomes large. Therefore, a distributed precoding scheme for CJT with minimal information exchange is required for practical implementation. Although pilot sequences in the uplink procedure were utilized to calculate the term related to the instantaneous CSI in \cite{jka2017} to reduce the amount of exchanged information for downlink precoding, the algorithm is still based on the iterative WMMSE method and the calculated term should be exchanged among different BSs per iteration, which is not acceptable yet.

In this paper, we propose a decentralized precoding scheme for CJT that allows the cooperating BSs to locally obtain near-optimal precoders with a minimal information exchange. In particular, our method does not require the instantaneous global CSI, which should be estimated and exchanged before precoding in each transmission block. Instead, we only need the statistical information of global channel (i.e., covariance matrices), which substantially reduces the signaling overhead in each transmission block. Specifically, we formulate the power minimization beamforming problem parameterized by the instantaneous global CSI, where precoders at all BSs are coupled by the interference constraints. To decouple the problem at individual BSs, we estimate the interference terms using the statistical channel information via deterministic equivalents (DE) \cite{RMT2011}. As such, the beamforming optimization is decentralized and can be solved by each BS individually.

\textbf{Notations:} We use lower-case letters, bold-face lower-case letters, bold-face upper-case letters, and math calligraphy letters to denote scalars, vectors, matrices, and sets, respectively. The transpose and conjugate transpose of a matrix $\mathbf{M}$ are denoted as $\mathbf{M}^{T}$ and $\mathbf{M}^{H}$, respectively. Besides, we denote the complex Gaussian distribution with mean $\mu$ and variance $\sigma^2$ as $\mathcal{C} \mathcal{N}(\mu, \sigma^2)$, the set $\mathcal{A}$ without the element $i$ as $\mathcal{A} \setminus i$, and the set $\mathcal{A}$ without the subset $\mathcal{A}_{i}$ as $\mathcal{A} \setminus \mathcal{A}_{i}$.

\section{Problem Formulation}
\label{sec:Model}
\subsection{System Model}
We consider the downlink of a multicell multiuser MIMO system with $N_B$ BSs serving $N_c$ UEs. Each BS has $N_T$ transmit antennas
and each UE is equipped with a single receive antenna. The collection of all UEs is denoted by $\mathcal{U} \triangleq \{1,2,\cdots, N_c\}$ and the set of BSs is represented by $\mathcal{T} \triangleq \{1,2,\cdots,N_B\}$. The subset $\mathcal{U}_p \subseteq \mathcal{U}$ with cardinality $U_p$ denotes the sets of UEs associated with BS $p$. For signal quality enhancement, each UE $i$ can be served by a group of cooperating BSs, denoted by $\mathcal{T}_i \subseteq \mathcal{T}$ with cardinality $T_i$.
Let $\mathbf{h}_{ip} \in \mathbb{C}^{N_T}$ be the channel vector from BS $p$ to UE $i$ and $\mathbf{w}_{ip} \in \mathbb{C}^{N_T}$ as the precoding vector of UE $i$ at the intended BS $p$. In particular, the channel vector from BS $p$ to UE $i$ is represented as $\mathbf{h}_{ip} = \mathbf{\Theta}_{ip}^{1/2} \mathbf{z}_{ip}$ where 
$\mathbf{z}_{ip}$  signifies the small-scale fading and comprises i.i.d complex Gaussian entries with zero mean and unit variance. The matrix 
$\mathbf{\Theta}_{ip} \in \mathbb{C}^{N_T \times N_T}$ denotes the covariance matrix of $\mathbf{h}_{ip}$, and the impact of pathloss stemming from
large-scale fading is inherently incorporated within it. The received signal 
at UE $i$, denoted as $y_i \in \mathbb{C}$, is a composite of the desired signal, intra-cell and inter-cell interference, which can be written as follows:
\begin{equation}  \label{eq:model}
  \begin{aligned}
    y_i  & = \sum_{p \in \mathcal{T}_i} \mathbf{h}_{ip}^H \mathbf{w}_{ip} s_i + \sum_{p \in \mathcal{T}_i} \mathbf{h}_{ip}^H \sum_{j \in \mathcal{U}_p \setminus i} \mathbf{w}_{jp} s_j \\
         & + \sum_{p \in \mathcal{T} \setminus \mathcal{T}_i} \mathbf{h}_{ip}^H \sum_{j \in \mathcal{U}_p } \mathbf{w}_{jp} s_j + n_i.
  \end{aligned}
\end{equation}
Note that signal $s_i$ signifies the data symbol intended to UE $i$ with zero mean and unit variance, and $n_i \sim \mathcal{C}\mathcal{N}(0,\sigma^2)$ denotes the additive white Gaussian noise. We further define the signal-to-interference-plus-noise ratio (SINR) at UE $i$ from BS $p$ as follows:
\begin{equation}  \label{eq:SINR}
  \Gamma_{ip} = \frac{|\mathbf{h}_{ip}^H \mathbf{w}_{ip}|^2}{\displaystyle \sum_{q \in \mathcal{T}} \sum_{j \in \mathcal{U}_{q} \setminus i} |\mathbf{h}_{iq}^H \mathbf{w}_{jq}|^2 + \sigma^2}.
\end{equation}
Note that this is an approximation of the original SINR at UE $i$ in (\ref{eq:originalSINR}) so that the fully decentralized precoding scheme can be realized. However, the original SINR formulation is still used for performance evaluation for fair comparison.
\begin{equation}  \label{eq:originalSINR}
  \Gamma_{i} = \frac{\Big|\sum_{p\in\mathcal{T}_{i}}\mathbf{h}_{ip}^H \mathbf{w}_{ip}\Big|^2}{\displaystyle \sum_{j \in \mathcal{U} \setminus i} \Big|\sum_{q \in \mathcal{T}_{j}} \mathbf{h}_{iq}^H \mathbf{w}_{jq}\Big|^2 + \sigma^2}.
\end{equation}

\subsection{Problem Formulation}
In the CoMP downlink, the BSs design precoders to jointly minimize the total transmit power subject to minimum SINR requirements from each BS to its serving UEs, denoted as $\gamma_{ip} \; \forall p, \forall i \in \mathcal{U}_p$.
This problem can be formulated as follows:
\begin{equation}  \label{eq:opt1}
  \begin{aligned}
    & \min_{\{\mathbf{w}_{ip}\}}  \sum_{p \in \mathcal{T}} \sum_{i \in \mathcal{U}_p} \Vert \mathbf{w}_{ip} \Vert_2^2 \\
    &\ \ \text{s.t.} \quad \Gamma_{ip} \geq \gamma_{ip}, \quad \forall p \in \mathcal{T}, \forall i \in \mathcal{U}_p.
  \end{aligned}
\end{equation}
Denote the inter-cell interference terms as $\tau_{iq}$ and $\epsilon_{iq}$, which represents the interference from BS $q$ that serves UE $i$ or not, respectively. The 
optimization problem in (\ref{eq:opt1}) can thus be reformulated as follows:
\begin{align}  \label{eq:opt2}
 \begin{aligned}
      & \min_{\{\mathbf{w}_{ip}, \epsilon_{iq}, \tau_{iq} \}}  
      \sum_{p \in \mathcal{T}} \sum_{i \in \mathcal{U}_p} \Vert \mathbf{w}_{ip} \Vert_2^2 \\
      &\ \text{s.t.}\ \frac{|\mathbf{h}_{ip}^H \mathbf{w}_{ip}|^2}{\displaystyle \sum_{j \in \mathcal{U}_p \setminus i}|\mathbf{h}_{ip}^H \mathbf{w}_{jp}|^2 +\!\sum_{q \in \mathcal{T}_i \setminus p} \tau_{iq} + \sum_{q \in \mathcal{T} \setminus  \mathcal{T}_i} \epsilon_{iq} + \sigma^2} \geq \gamma_{ip}, \\ 
      &\quad \quad \quad \quad \quad \quad \quad \quad \quad \quad \quad \quad \quad \quad \quad \quad \quad \forall p \in \mathcal{T}, \forall i \in \mathcal{U}_p,\\
      & \quad \quad \sum_{j \in \mathcal{U}_q \setminus i } |\mathbf{h}_{iq}^H \mathbf{w}_{jq}|^2 \leq \tau_{iq}, \quad \forall q, \forall i \in \mathcal{U}_q, \\
      & \quad \quad \sum_{j \in \mathcal{U}_q} |\mathbf{h}_{iq}^H \mathbf{w}_{jq}|^2 \leq \epsilon_{iq}, \quad \forall q, \forall i \notin \mathcal{U}_q.
  \end{aligned}
\end{align}
Following the similar analysis in \cite{ato2011}, it can be established that (\ref{eq:opt1}) and (\ref{eq:opt2}) are equivalently at the optimal solution, where the interference constraints in (\ref{eq:opt2})
are satisfied with equality. Therefore, if $\{ \tau_{iq}\}$'s and $\{ \epsilon_{iq}\}$'s are fixed, we can decouple Problem (\ref{eq:opt2}) for the $p$-th BS, $\forall p \in \mathcal{T}$ as follows:
\begin{align}  \label{eq:opt3}
  \begin{aligned}
      & \min_{\{\mathbf{w}_{ip} \}}  
       \sum_{i \in \mathcal{U}_p} \Vert \mathbf{w}_{ip} \Vert_2^2 \\
      &\ \text{s.t.} \ \frac{|\mathbf{h}_{ip}^H \mathbf{w}_{ip}|^2}{\displaystyle \sum_{j \in \mathcal{U}_p \setminus i} |\mathbf{h}_{ip}^H \mathbf{w}_{jp}|^2 + \sum_{q \in \mathcal{T}_i \setminus p} \tau_{iq} + \sum_{q \in \mathcal{T} \setminus  \mathcal{T}_i} \epsilon_{iq} + \sigma^2} \geq \gamma_{ip}, \\
      &\quad \quad \quad \quad \quad \quad \quad \quad \quad \quad \quad \quad \quad \quad \quad \quad \quad \quad \quad \quad \quad \forall i \in \mathcal{U}_p,\\
      &\ \ \quad \sum_{j \in \mathcal{U}_p \setminus i } |\mathbf{h}_{ip}^H \mathbf{w}_{jp}|^2 \leq \tau_{ip}, \quad \forall i \in \mathcal{U}_p, \\
      &\ \ \quad \sum_{j \in \mathcal{U}_p} |\mathbf{h}_{ip}^H \mathbf{w}_{jp}|^2 \leq \epsilon_{ip}, \quad \forall i \notin \mathcal{U}_p. 
  \end{aligned}
\end{align}
The sub-problem for each BS can be easily transformed to a convex problem \cite{zql2006} like second order cone programming (SOCP) and semidefinite programming (SDP) , which can be solved with many off-the-shelf solvers.

\section{Decentralized CJT Precoding}
As mentioned in the previous section, once the inter-cell interference terms are fixed, the centralized precoding problem can be solved in a decentralized form. There remains a question: How can we accurately estimate the interference with minimal information exchange? In this section, we propose to utilize DE, a powerful tool that can calculate the values of certain functions of large random matrices, to estimate the interference.

\subsection{Centralized Calculation of Interference Terms}
First of all, without considering the cost of information exchange, we provide a computation method for the interference terms via the uplink-downlink duality-based centralized optimal precoding approach \cite{hda2010}, which was originally proposed for multi-cell non-CJT systems. Following the Lagrangian analysis similar to \cite{has2019}, the inter-cell interference terms are expressed respectively as
\begin{equation}  \label{eq:tau}
  \tau_{iq} = \sum_{j \in \mathcal{U}_q \setminus i } \frac{1}{N_T}  \delta_{jq} |\mathbf{h}_{iq}^H \hat{\mathbf{w}}_{jq}|^2, \quad \forall q, \forall i \in \mathcal{U}_q,
\end{equation}
\begin{equation}  \label{eq:eps}
  \epsilon_{iq} = \sum_{j \in \mathcal{U}_q } \frac{1}{N_T} \delta_{jq} |\mathbf{h}_{iq}^H \hat{\mathbf{w}}_{jq}|^2, \quad \forall q, \forall i \notin \mathcal{U}_q,
\end{equation}
where $\delta_{ip}$ is the scaling factor that connects $\mathbf{w}_{ip}$ with $\hat{\mathbf{w}}_{ip}$. Therefore, the optimal precoding vector is $\mathbf{w}_{ip} = \sqrt{\frac{\delta_{ip}}{N_T}} \hat{\mathbf{w}}_{ip}$, where
\begin{equation}
  \hat{\mathbf{w}}_{ip} = (N_T \mathbf{I} +  \sum_{j \in \mathcal{U} \setminus i} \sum_{q \in \mathcal{T}_j} \lambda_{jq}^* \mathbf{h}_{jp} \mathbf{h}_{jp}^H)^{-1} \mathbf{h}_{ip}
\end{equation}
with $\lambda_{ip}$ being the unique fixed point solution of 
\begin{equation}
  \lambda_{ip} = \frac{\gamma_{ip}}{\mathbf{h}_{ip}^H(N_T \mathbf{I} +  \sum_{j \in \mathcal{U} \setminus i} \sum_{q \in \mathcal{T}_j} \lambda_{jq} \mathbf{h}_{jp} \mathbf{h}_{jp}^H)^{-1} \mathbf{h}_{ip}},
\end{equation} 
and the scaling factors $\{\delta_{ip}\}$'s are determined by $\boldsymbol{\delta} = N_T \sigma^2 \mathbf{F}^{-1} \boldsymbol{1}$ with the
definitions $\boldsymbol{\delta}_p = \{{\delta}_{ip}\}_{i \in \mathcal{U}_p}$ and $\boldsymbol{\delta} = [\boldsymbol{\delta}_1, \boldsymbol{\delta}_2, \cdots, \boldsymbol{\delta}_p]^T$. Here,
$\boldsymbol{1}$ is the $(\sum_{p \in \mathcal{T}} U_p) $-dimensional all ones-vector and the shape of $\mathbf{F}$ is as follows:
\begin{equation} \label{eq:F1}
  \mathbf{F}=\left[\begin{array}{cccc}
      \mathbf{F}^{11} & \mathbf{F}^{12} & \ldots & \mathbf{F}^{1 N_B} \\
      \mathbf{F}^{21} & \mathbf{F}^{22} & \ldots & \mathbf{F}^{2 N_B} \\
      \cdot & & & \\
      \cdot & & & \\
      \cdot & & & \\
      \mathbf{F}^{N_B 1} & \mathbf{F}^{N_B 2} & \ldots & \mathbf{F}^{N_B N_B}
      \end{array}\right],
\end{equation}
where the $(i,j)^{\text{th}}$ element of the so-called coupling matrix $\mathbf{F}^{pq}$ is
\begin{equation} \label{eq:F2}
    {F}_{{ij}}^{{pq}}=\left\{\begin{array}{ll}
        \frac{1}{\gamma_{ip}}\left|\hat{\mathbf{w}}_{ip}^{H} \mathbf{h}_{ip}\right|^{2}, & \text { if } q=p \text { and } j=i, \\
        -\left|\hat{\mathbf{w}}_{jq}^{H} \mathbf{h}_{iq}\right|^{2}, & \text { if } q \in \mathcal{T} \text { and } j \in \mathcal{U}_q \setminus i,\\
        0, & \text { else } (q \in \mathcal{T} \setminus p, j = i).
        \end{array}\right.
\end{equation}

However, such a method for computing $\{\epsilon_{iq}\}$'s and $\{\tau_{iq}\}$'s necessitates inter-base station communication for exchange of channel vectors, which incurs substantial costs of communication and storage. With the DE techniques from random matrix theories, we can characterize the asymptotic behavior of interference terms by leveraging the statistical channel information without knowing every entries of channel vector exactly. Then we can proffer a set of good
approximations for the interference terms by utilizing only the covariance matrices of global CSI. 



\subsection{Approximation of Interference Terms}  \label{sec:LSA}
To utilize the DE theory, it is essential to establish the following judicious assumptions concerning system dimensions,
\begin{assumption}   \label{eq:ass2}
    $0<\lim \limits_{ N_T \to \infty} {\inf} \frac{N_c}{N_T} \leq \lim \limits_{ N_T \to \infty} {\sup} \frac{N_c}{N_T} < \infty$.
\end{assumption}
\begin{assumption}   \label{eq:ass3}
   The spectrum norm of $\boldsymbol{\Theta}_{ip}$ is uniformly bounded, i.e. $\lim \limits_{ N_T \to \infty} \sup \max _{\forall i, p}\left\{\left\|\Theta_{ip}\right\|\right\}<\infty$. 
\end{assumption}

Firstly, we derive the DE of Lagrangian multipliers $\{\lambda_{ip}\}$. By employing the analogous analysis involving the rank-1 perturbation lemma \cite{SJW1995} and the trace lemma \cite{bai1998}, we obtain the following results.
\begin{theorem} \label{eq:th1}
   Let Assumptions (\ref{eq:ass2}) and (\ref{eq:ass3}) hold. We have $\max_{i,p} |\lambda_{ip}^* - \overline{\lambda}_{ip}| \to 0$ almost surely where
   \begin{equation} \label{eq:delamda}
     \overline{\lambda}_{ip} = \frac{\gamma_{ip}}{\overline{m}_{ip}} \quad \forall p \in \mathcal{T}, i \in \mathcal{U}_p.
   \end{equation}
   The value of $\overline{m}_{ip}$ is determined as the unique non-negative solution to the following system of equations, computed for 
   $p \in \mathcal{T}, i \in \mathcal{U}$, 
   \begin{equation}  \label{eq:dem}
    \bar{m}_{ip} = \operatorname{Tr}(\boldsymbol{\Theta}_{ip} (\sum_{j \in \mathcal{U}} \frac{(\sum_{q \in \mathcal{T}_j } \lambda_{jq} ) \boldsymbol{\Theta}_{jp}}{1+ (\sum_{q \in \mathcal{T}_j } \lambda_{jq} ) \bar{m}_{jp}} + N_T \mathbf{I})^{-1}).
   \end{equation}
\end{theorem}
Theorem \ref{eq:th1} establishes that in the large system limitation, the asymptotic value of $\lambda_{ip}$ can be decoupled from the specific attributes of channel vectors. 
Specifically, it can be exclusively computed by leveraging the covariance matrix.

Then, we need to obtain the DE of entries in coupling matrix $\mathbf{F}$ and the scaling factor $\boldsymbol{\delta}$. Therefore, the deterministic approximations of $\{\lambda_{ip}^*\}$'s in Theorem \ref{eq:th1} are applied to calculate $\hat{\mathbf{w}}_{ip}$ asymptotically as $\hat{\overline{\mathbf{w}}}_{ip}= (N_T \mathbf{I} +  \sum_{j \in \mathcal{U} \setminus i} 
\sum_{q \in \mathcal{T}_j} \overline{\lambda}_{jq} \mathbf{h}_{jp} \mathbf{h}_{jp}^H)^{-1} \mathbf{h}_{ip}$ in the dual problem. Taking
$\{\hat{\overline{\mathbf{w}}}_{ip}\}$ to the coupling matrix, we have the following result.
\begin{theorem} \label{eq:th2}
  Let Assumptions  (\ref{eq:ass2}) and (\ref{eq:ass3}) hold. Then given the set of $\overline{\lambda}_{ip}$ and 
  $\overline{m}_{ip}$, $\forall p \in \mathcal{T}, \forall i \in \mathcal{U}_p$ in Theorem \ref{eq:th1}, we have $|F^{pq}_{ij} - \overline{F}^{pq}_{ij}| \to 0$
  almost surely with 
  \begin{equation}  \label{eq:deF}
    \overline{F}_{i j}^{pq}=\left\{\begin{array}{ll}
        \frac{1}{\gamma_{i p}} \bar{m}_{ip}^2, & \text { if } q=p \text { and } j=i, \\
        -\frac{1}{N_T} \frac{\bar{m}_{j,i,q}^{\prime}}{(1 + (\sum_{r \in \mathcal{T}_i} \bar{\lambda}_{ir}) \bar{m}_{iq})^2}, & \text { if } q \in \mathcal{T} \text { and } j \in \mathcal{U}_{q} \backslash i, \\
        0, & \text { else }(q \in \mathcal{T} \backslash p, j=i),
        \end{array}\right.
\end{equation}
where we have $[\bar{m}_{1,i,q}^{\prime},\bar{m}_{2,i,q}^{\prime},\cdots,\bar{m}_{N_c,i,q}^{\prime}]^T= (\mathbf{I}_{N_c} - \mathbf{L}_q)^{-1} \mathbf{u}_{iq}$, $\forall 
p \in \mathcal{T}, \forall i \in \mathcal{U}_p$, with entries of $\mathbf{L}_q$ being
\begin{equation}
  [L_q]_{hl} = \frac{1}{N^2} \frac{\operatorname{Tr} (\boldsymbol{\Theta}_{hq} \mathbf{T}_q (\sum_{r \in \mathcal{T}_l} \lambda_{rl})^2 \boldsymbol{\Theta}_{lq} \mathbf{T}_q)}{[1 + (\sum_{r \in \mathcal{T}_l} \lambda_{rl}) \bar{m}_{lq}]^2}.
\end{equation}
Besides,
\begin{equation}
  \begin{aligned}
    \mathbf{u}_{iq} =  \Big[\frac{1}{N} \operatorname{Tr} (\boldsymbol{\Theta}_{1q} \mathbf{T}_q \boldsymbol{\Theta}_{iq} \mathbf{T}_q), & \frac{1}{N} \operatorname{Tr} (\boldsymbol{\Theta}_{2q} \mathbf{T}_q \boldsymbol{\Theta}_{iq} \mathbf{T}_q),\cdots,  \\ 
    & \frac{1}{N} \operatorname{Tr} (\boldsymbol{\Theta}_{N_c q} \mathbf{T}_q \boldsymbol{\Theta}_{iq} \mathbf{T}_q)\Big]^T
  \end{aligned}
\end{equation}
with $\mathbf{T}_q=(\frac{1}{N} \sum_{k \in \mathcal{U}} \frac{(\sum_{r \in \mathcal{T}_k} \overline{\lambda}_{kr}) \boldsymbol{\Theta}_{kq}}{1 + (\sum_{r \in \mathcal{T}_k } \overline{\lambda}_{kr}) \overline{m}_{kq}} + \mathbf{I})^{-1}$.
The DE of $\{\overline{F}_{ij}^{pq}\}$ can be employed to compute the asymptotically optimal scaling factors as $\overline{\boldsymbol{\delta}} = N_T \sigma^2 \overline{\boldsymbol{F}}^{-1} \boldsymbol{1}$.  
\end{theorem}
Based on these theorems, we get all DE of all scalar parameters related to the interference terms. Then, by plugging the DE of $\boldsymbol{\delta}$ and $F_{ij}^{pq}$ into (\ref{eq:tau}) and (\ref{eq:eps}), we get the approximation of interference terms as $\bar{\tau}_{iq} = - \frac{1}{N} \sum_{j \in \mathcal{U}_q \setminus i }  \overline{\delta}_{jq} \bar{F}^{qq}_{ij},  \forall q, \forall i \in \mathcal{U}_q$ and $\bar{\epsilon}_{iq} = -\frac{1}{N} \sum_{j \in \mathcal{U}_q}  \overline{\delta}_{jq} \bar{F}^{qq}_{ij}  ,\forall q, \forall i \notin \mathcal{U}_q$. It is notable that the approximation necessitates solely the exchange of channel covariance matrices $\{\boldsymbol{\Theta}_{ip}\}$'s, which are slow variables and can be exchanged only once for a long time.

\section{Simulation Results}
We simulate a multi-cell downlink celluar network with $N_B=3$ BSs and $N_c=20$ UEs in total, and the UEs are uniformly distributed in the coverage area of BSs. In particular, BS1, BS2, and BS3 serve the UE1-10, UE6-15, and UE11-20, respectively, which means that UE6-15 are served by two BSs simutaneously. The channel vectors $\{\mathbf{h}_{ip}\}$ are generated from Quasi Deterministic Radio Channel Generator (QuaDRiGa) \cite{fbu2014}, which is calibrated against 3rd Generation Partnership Project (3GPP) channel models. The noise variance is given by $\sigma^2=10^{\frac{1}{U_p} \sum_{p \in \mathcal{T}}\sum_{i \in \mathcal{U}_p} \log _{10}\left\|\mathbf{h}_{ip}\right\|_2^2} \times 10^{-\frac{\mathrm{SNR}}{10}}$, where $\mathrm{SNR} = 20$ dB is the average receive SNR for all users when no precoding is applied. To solve the sub-problem at each BS, we use the SOCP solver from the CVX toolbox \cite{cvx}. Our simulation results are averaged over 100 randomly generated channel realizations.

\begin{figure}[t]
\centering
\includegraphics[width=3in]{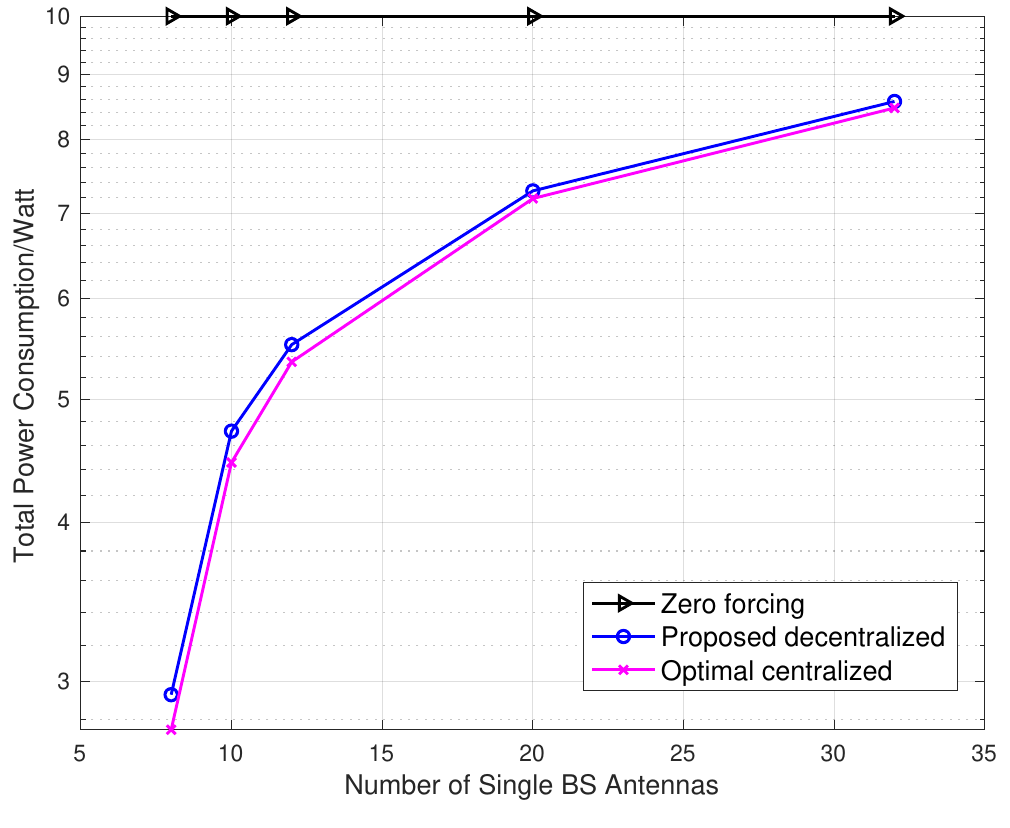}
\caption{Total power consumption versus the number of antennas at each BS.}
\label{powerfig}
\end{figure}

We evaluate the transmit power used of all BSs versus the number of antennas at each BS in Fig. \ref{powerfig}. The centralized ZF precoder is considered as a baseline, where the instantaneous CSI from all BSs are collected in a central unit to perform the ZF operation. Besides, a lower bound with the uplink-downlink duality-based optimal centralized precoder, i.e., $\{\mathbf{w}_{ip}\}$ in Section 3.1, is also exhibited. In particular, for fair comparison, the ZF precoder is normalized to $10$ Watt in total power, and the SINR constraints used for our proposed decentralized and the uplink-downlink duality-based centralized precoders are generated from the normalized ZF precoder, so that the sum rate of all schemes are the same. Compared with the ZF precoder, the proposed decentralized one achieves much lower power consumption, and is extremely close to the lower bound, which only consumes $2 \sim 7\%$ more power. This demonstrates the accuracy of DE for approximating the true interference from the optimal centralized precoding scheme using the covariance matrices of global CSI and also shows the effectiveness of our proposed decentralized precoding scheme for CJT with only a little exchanged information.
\vspace{-0.2cm}
\section{Conclusions and Discussions}
In this paper, we proposed a decentralized framework for power minimization in coherent joint transmission. The key idea is to provide robust approximations for the interference terms by deterministic equivalents, solely based on statistical channel information. Subsequently, we obtain the precoding vectors by independently solving BS-based sub-problems. The power consumption of the proposed distributed precoder is close to that of the optimal centralized precoder. In the future work, we will extend the precoding scheme to the multi-stream scenario and investigate more low-complexity solvers for solving BS-based sub-problems.



\end{document}